\documentclass[%
nofootinbib,
 amsmath,amssymb,
 aps,
]{revtex4-1}

\usepackage{graphicx}   
\usepackage{verbatim}   
\usepackage{color}      
\usepackage{subfigure}  
\usepackage{hyperref}   
\usepackage{soul}
\usepackage{hyperref}
\usepackage{tabularx}

\usepackage{float}
\usepackage{amssymb}
\usepackage{mathtools}
\usepackage{setspace}
\usepackage{amsmath}
\usepackage{bm}
\usepackage{outlines}
\usepackage{xcolor,cancel}

\usepackage[utf8]{inputenc}
\usepackage[T1]{fontenc}
\usepackage{mathptmx}
\usepackage{lmodern} 

\usepackage{dcolumn}
\usepackage{float}
\usepackage{tcolorbox}
\usepackage{graphicx}
\usepackage{multirow}

\def\exp{{\rm {exp}}}
\def\ln{{\rm {ln}}}
\def\Ca{{\it {Ca}}}

\def\Re{{{\it Re}}}

\def\Pr{{{\it Pr}}}
\def\Sc{{{\it Sc}}}
\def\We{{{\it We}}}
\def\Ja{{{\it Ja}}}
\def\Le{{{\it Le}}}
\def\Sh{{{\it Sh}}}
\def\Nu{{{\it Nu}}}
\def\ln{{\rm ln}}

\begin{document}

\preprint{APS/123-QED}

\title{Dynamics of an evaporating drop migrating in a Poiseuille flow}

\author{Anubhav Dubey} 
\affiliation{%
Department of Mechanical Engineering, Indian Institute of Technology Kanpur, Kanpur- 208016, Uttar Pradesh, India}

\author{Kirti Chandra Sahu}
\affiliation{%
Department of Chemical Engineering, Indian Institute of Technology Hyderabad, Sangareddy, Telangana, 502 285, India
}%

\author{Gautam Biswas}
\affiliation{%
Department of Mechanical Engineering, Indian Institute of Technology Kanpur, Kanpur 208016, Uttar Pradesh, India}

\date{\today}

\begin{abstract}
The evaporation of a liquid drop of initial diameter ($D_{drop}$) migrating in a tube of diameter ($D_0$) is investigated using the coupled level set and volume of fluid (CLSVOF) method focusing on determining the heat and mass transfer coefficients for a deforming drop. A robust phase change model is developed using an embedded boundary method under a finite difference framework to handle vaporizing flows. The model is extensively validated through simulations of benchmark problems such as arbitrary evaporation of a static drop and reproduction of psychrometric data. The results show that the Sherwood number ($\Sh$) and the Nusselt number ($\Nu$) reach a steady value after an initial transient period for the drop subjected to Hagen-Poiseuille flow. A parametric study is conducted to investigate the effect of drop deformation on the rate of evaporation. It is observed that Stefan flow due to evaporation has a negligible impact on the drop deformation dynamics. We also observed that, for different values of $D_{drop}/D_0$, the $\Sh$ follows a linear correlation with $\Re^{1/2}\Sc^{1/3}$.
\end{abstract}

\maketitle 

\section{Introduction} \label{sec:intro}
The evaporation of liquid droplets in a gaseous environment plays a vital role in various industrial and natural processes, including cooling towers, inkjet printing, desiccation, biomedical applications \cite{Dugas2005Droplet}, spray combustion \cite{sazhin2006advanced}, and the formation of falling raindrops, fog, and clouds \cite{Pruppacher1998Microphysics,tripathi2015evaporating}. The review of early studies on drop evaporation, ignition, and combustion can be found in Refs. \cite{ayyaswamy1995direct,ayyaswamy1996mathematical}.

For a spherical drop, Spalding \cite{spalding1955some} proposed an analytical solution for the temporal evolution of drop diameter, commonly referred to as the $d^2$ law of evaporation, by considering concentration and temperature gradients individually as driving potentials for evaporation. Ranz and Marshall \cite{rantz1952evaporation} emphasized the importance of forced convection of the surrounding medium and provided independent correlations for heat and mass transfer rates of pure liquid drops and water drops containing dissolved and suspended solids. For a falling drop undergoing evaporation, Huang and Ayyaswamy \cite{Huang1990Evaporation} studied the quasi-steady evaporation of an n-heptane drop moving in the intermediate Reynolds number regime $(Re = O(100))$, while Jog, Ayyaswamy, and Cohen \cite{Jog1997Evaporation} introduced a novel method that included the formulation for continuous phase modulation by imposing slow translator-motion of the drop as a perturbation to uniform radial flow associated with vigorous evaporation at the drop-surface. Tanguy et al. \cite{tanguy2007level} and Schlottke and Weigand \cite{schlottke2008direct} performed numerical simulations to investigate the dynamics of an evaporating drop at various temperatures. Tripathi and Sahu \cite{tripathi2015evaporating} performed three-dimensional numerical simulations in the volume-of-fluid framework to study the dynamics of a falling drop undergoing evaporation. They \cite{schlottke2008direct,tripathi2015evaporating} analysed the spatio-temporal evolution of the drop shape and the contours of the vapour concentration generated due to evaporation. Lupo et al. \cite{lupo2019immersed} investigated the evaporation of a drop in laminar and turbulent gas flows. 

Previous studies have emphasized the necessity of employing a comprehensive Navier-Stokes equation solution by employing different computational methods, such as the volume-of-fluid (VOF), diffuse-interface method, combination of the level set method and volume of fluid (CLSVOF) \cite{pal2023clsvof} approaches \cite{schlottke2008direct, scapin2020volume, tripathi2015evaporating,pal2023clsvof} to model the dynamics of drop evaporation accurately. Additionally, a finite-volume-based method with adaptive mesh refinement, utilizing the Basilisk flow solver \cite{zhao2022boiling}, has been employed to investigate boiling and evaporation models for liquid-gas flows. This method has been applied to study phenomena such as levitating drops over heated solid surfaces and liquid pools. However, despite these notable efforts, the evaporation of liquid drops near walls remains a relatively unexplored area, which is the focus of the present investigation.
 
In the case of droplets migrating within channels or tubes, the boundary layer surrounding the droplets governs the heat and mass transport rates. Some researchers have investigated the dynamics of a drop migrating through a pipe or channel without undergoing evaporation, which is reviewed below. Nath et al. \cite{nath2017migration} conducted a comprehensive study of drop dynamics in a microcapillary using the coupled level set and volume of fluid (CLSVOF) method. Their research demonstrated that a liquid drop with neutral buoyancy could deform and disintegrate based on the balance between viscous and interfacial tension forces. Ho and Leal \cite{Ho1975CreepingMotion} examined the drop deformation experimentally to assess the significance of the viscosity ratio and the capillary number. The critical capillary number values that could cause drop disintegration for a particular viscosity ratio and drop size were identified. A few researchers have also investigated the dynamics of a compound drop moving inside a circular tube theoretically \cite{song2010stokes} and by conducting numerical simulations \cite{borthakur2018dynamics,che2018flow}. Furthermore, the morphological changes of a compound drop as it moves through a gradually constricted tube with the inhomogeneous flow were studied \cite{zhou2008deformation}. Using the lattice Boltzmann method, Liu et al. \cite{liu2021deformation} investigated the dynamics of a compound drop under three-dimensional oscillatory shear flow. They demonstrated that the inner drop might rotate counterintuitively in a direction opposite to the outer one due to high pressure near two tips inside the outer drop. The compound drop undergoes more significant deformation when either drop is less viscous, decreasing the synchronization between the inner and outer drops. The abovementioned brief review indicates that although many researchers have examined the dynamics of a drop moving in a constrained geometry, no one has done so in the presence of evaporation, despite it being significantly relevant in many real-world applications.

As the brief review indicates, previous research on forced convective evaporation has focused on the evaporation of liquid droplets far from the channel boundary. In the present study, we examine the influence of a nearby wall on the heat and mass transfer dynamics of an evaporating droplet, which is commonly encountered in practical applications. We consider that the droplet evaporates while migrating inside a tube, exposing it to a parabolic velocity profile. Specifically, we focus on determining the heat and mass transfer coefficients for a deforming drop. We investigate the effect of the Weber number, which determines the relative importance of aerodynamic drag over the interfacial tension forces. Our study reveals the effect of drop deformation on the evaporation rate, which varies depending on the drop size/Weber number. Additionally, we analyze how the density ratio, viscosity ratio, and degree of superheat affect the evaporation dynamics of the drop within the tube. Our results highlight the relative importance of convective gas flow and the Stefan flow associated with evaporation. We vary initial diameter droplet $(D_{drop})$, characterized by the aspect ratio $a = {D_{drop}/D_0}$, while keeping the tube diameter $(D_0=2R_0)$ constant. Our findings demonstrate that, for different initial droplet sizes, the Sherwood number $(Sh)$, which is the ratio of the convection mass transfer to the rate of diffusion mass transport and indicates non-dimensional mass transfer gradient (mathematically defined in section \ref{sec:dis}), exhibits a linear relationship with $Re^{1/2}Sc^{1/3}$, where $\Re ~ (=\rho_{g} U_{avg} D_0 / \mu_{g})$ and $Sc ~ ( =\mu_g / \rho_g D_{\alpha g})$ represent the Reynolds number and Schmidt number of the vapor phase in the surrounding gas, respectively. However, the slope of the $Sh$ versus $Re^{1/2}Sc^{1/3}$ relationship depends on the aspect ratio ($a$). Here, $U_{avg}$ is the average velocity of the imposed pressure-driven flow; $D_{\alpha g}$ is the mass diffusion coefficient in the gaseous medium; $\rho_l,\rho_g$ and $\mu_l,\mu_g$ are the densities and viscosities of liquid and gas phases, respectively. 

The rest of the paper is organised as follows. The problem is formulated in section \ref{sec:form}, where the governing equations, numerical method and validation of the solver are discussed. The results are presented in section \ref{sec:dis}. In section \ref{sec:conc}, concluding remarks are provided.   

\section{Formulation} \label{sec:form}

\begin{figure}
\centering 
\includegraphics[width=0.45\textwidth]{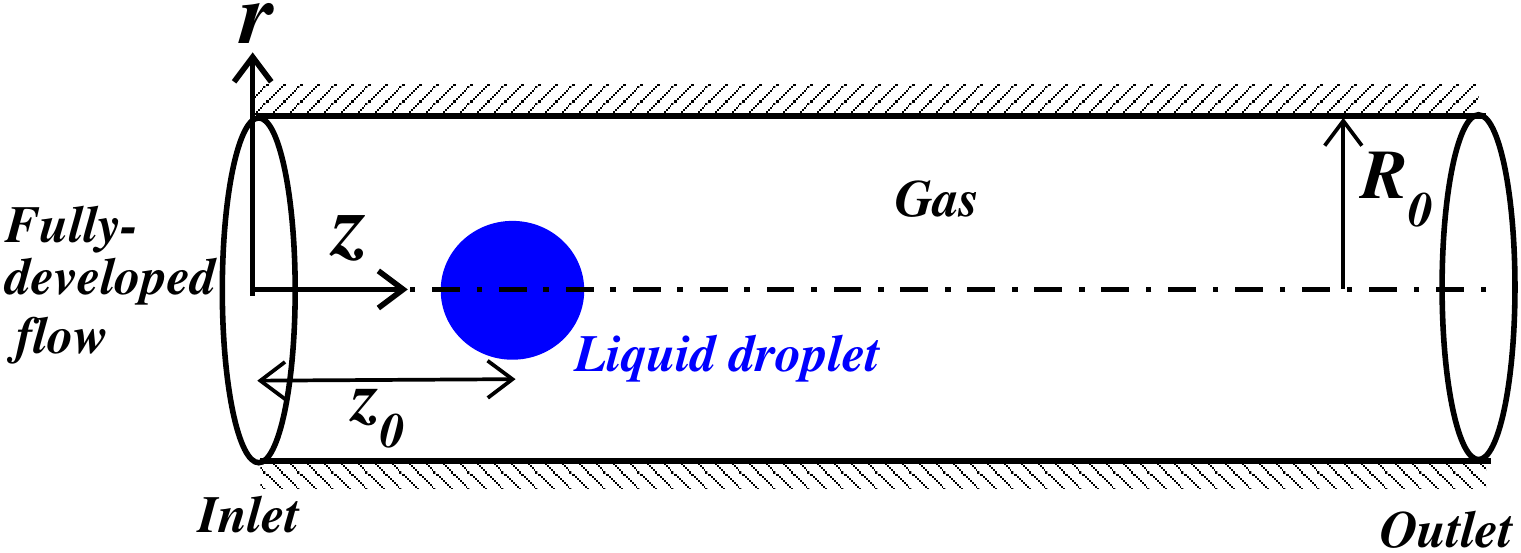} 
\caption{Schematic diagram of an initially spherical drop of radius $R$ migrating in a tube of radius $R_0$ while undergoing evaporation. The velocity profile at the inlet of the tube is assumed to be fully developed. The bulk liquid and the interfacial temperature are also assumed to be at saturation temperature.}
\label{fig1}
\end{figure}

We investigate the dynamics of an evaporating drop of initial diameter $D_{drop}$ migrating in a cylindrical tube of diameter $D_0$ and length, $L=10D_0$. A sufficiently long tube is considered to capture the transient evolution of drop. In an axisymmetric framework, the numerical simulations are performed using the CLSVOF algorithm within a finite-difference formulation. Figure \ref{fig1} depicts the schematic diagram showing the initial configuration. The fluids are assumed to be Newtonian and incompressible. The thermophysical properties of the fluids are assumed to be constant during the evaporation process. 
\subsection{Governing equations}
The drop migrating in a gaseous medium in a tube is governed by the continuity and momentum equations, which are given by
\begin{eqnarray}
\nabla \cdot{\vec V}  &=& 0,  \label{conti} \\
\rho (\phi) \left [ {\partial \vec{V} \over \partial t} + \vec V \cdot \nabla \vec V \right] &=& -\nabla P + \nabla \cdot \left [\mu (\phi) \left (\nabla \vec{V} + \nabla \vec{V}^T \right) \right]  \nonumber \\
&+& \sigma \kappa(\phi) \hat {n} \delta_s (\phi), \label{eq:momentum} 
\end{eqnarray}
where ${\vec V}(u,v)$ is the velocity field, wherein $u$ and $v$ represent the component of velocity in the axial $(z)$ and radial $(r)$ directions, respectively; $P$ denotes pressure field; $\rho (\phi)$ and $\mu (\phi)$ are the volume fraction-based density and viscosity, respectively; $\sigma$ represents the surface tension; $\hat {n} (= \nabla \phi / |\nabla \phi|)$ is the outward pointing unit normal vector at the interface, wherein $\phi$ represents the level-set function; $\kappa (= - \nabla \cdot \hat {n})$ is the mean curvature of the interface; $\delta_s (\phi)$ is a Dirac delta distribution function which is zero everywhere except at the interface; $t$ denotes time. The single-phase continuity equation (Eq. \ref{conti}) is valid in bulk fluid regions only, i.e., cells containing pure fluids. However, the interfacial region undergoes a velocity jump which can be accounted for in the model using an augmented continuity equation in the one-fluid formulation used in the present study. The source term for the modified continuity equation is calculated considering the mass transfer at the interface due to evaporation. In Eq. (\ref{eq:momentum}), the density and viscosity are defined based on the volume fraction of the liquid. The last term accounts for the effects of the surface tension force in the interfacial region. The surface tension is incorporated as a body force following the CSF model of Ref. \cite{brackbill1992continuum}.  

For incorporating evaporation dynamics, we need to solve energy as well as species transport equation, which are given by
\begin{eqnarray}
{\partial T \over \partial t} + \vec V \cdot \nabla T &=& \nabla \cdot \left( {k \over \rho  c_{p}} \nabla T \right), \label{eq:temp1} \\
{\partial Y \over \partial t} + \vec V \cdot \nabla Y &=& \nabla \cdot \left( D_{\alpha g} \nabla Y \right), 
\label{eq:temp2} 
\end{eqnarray}
where $T$ represents the temperature field, $Y$ represents the mass fraction field, and $k$ and $c_p$  are conductivity and specific heat, respectively. The energy equation is solved in both liquid and gaseous media. We assume that the liquid drop is mono-component and gas does not diffuse into the liquid drop. Therefore, Eq. (\ref{eq:temp2}) is solved only in the gaseous phase. 

\subsection{Jump conditions at interface}

The jump in the velocity field at the interface induced by the phenomena of mass transfer can be mathematically expressed as follows:
\begin{equation}
|| {(\vec{V}-\vec{V}_I)} || \cdot \hat {n} = \left( {1 \over \rho_g} - {1 \over \rho_l}\right) {\dot m}, \label{jump3}
\end{equation}
where $\vec{V}_I$ is the instantaneous interfacial velocity vector and $\rho_l$ and $\rho_g$ are the densities of liquid and gas phases respectively. Thus, the modified continuity equation in the one-fluid formulation can be written as \citep{Welch2000Volume, Tomar2005Numerical}
\begin{equation}
\int_{S_c} \vec{V}^{n+1} \cdot \hat {n} dS + \int_{S_I(t)} \left( {1 \over \rho_g} - {1 \over \rho_l}\right)  {\dot m} dS = 0. \label{jump4}
\end{equation}
Here, the first term in the above equation is the integral over the cell surface ${S_c}$, and the second term represents the integral over the interfacial area ${S_I(t)}$ in the two-phase cell under consideration. Similarly, the thermal energy jump at the interface, considering the mass of the liquid undergoing a phase change, can be expressed as \cite{irfan2017front}
\begin{equation}
{\dot m}  h_{lg}=  (q_g -q_l) \cdot \hat {n}, \label{jump2}
\end{equation}
where $h_{lg}$ is the latent heat of vaporization, $q_g$  is heat flowing from the vapor to the interface, and $q_l$ is the heat flowing from the interface towards the liquid.

The vapor concentration field is also discontinuous at the interface. Under the assumptions of mono-component liquid drop and no diffusion of gas into the liquid, mass balance at the interface results in the following equation for vaporization rate \cite{irfan2017front}:
\begin{equation}
{\dot m}  = {{ \rho_g D_{\alpha g} \nabla Y \cdot  \hat {n}} \over (1 - Y_\Gamma)}, \label{eq8}
\end{equation}
where $Y_\Gamma$  is the species concentration at the interface.

It is to be noted that, unlike boiling, the temperature at the interface in the case of evaporation need not be at a saturation temperature. It is coupled with the interface mass fraction through the Clausius-Clapeyron equation, which is given by \cite{irfan2017front}
\begin{equation}
P_{vap}^\Gamma  = P_{atm} \exp \left ( - {h_{lg} M_{vap} \over R} \left( {1 \over T_\Gamma} - {1 \over T_{sat}} \right) \right),
\end{equation}
where $P_{atm}$ is the ambient pressure, $M_g$ and $M_{vap}$ denote the molar masses for surrounding gas and the liquid vapor, respectively, $T_\Gamma$ represent the interface temperature, $T_{sat}$ is saturation temperature corresponding to given ambient pressure, $P_{vap}^\Gamma$  denote saturation vapor pressure at the interface and $Y_\Gamma$ is the mass fraction corresponding to $P_{vap}^\Gamma$.

\subsection{Capturing the interface dynamics}

The interfacial dynamics are modeled using the level-set function $(\phi)$ and the volume fraction of the liquid phase ($F$) by solving the following advection equations \cite{Gerlach2006Comparison}:
\begin{eqnarray}
{\partial \phi \over \partial t}+ \vec {V} \cdot \nabla  \phi &=& 0, \label{eq:ls} \\
{\partial F \over \partial t} + \vec {V} \cdot \nabla  F  &=& 0, \label{eq:ls2}
\label{eq:clsvof}
\end{eqnarray}
where $F=0$ for the gaseous phase and $F=1$ for the liquid phase. For a cell containing both the liquid and gaseous phases, the value of $F$ lies between 0 and 1. Similarly, the values of the level-set function $(\phi)$ for the gaseous and liquid phases are $\epsilon$ and $-\epsilon$, respectively, and 0 at the interface separating the fluids. Here, $\epsilon$ is the numerical thickness of the interface. In the present study, we used $\epsilon=1.5\Delta$, where $\Delta$ is the grid size.

The density and dynamic viscosity of both phases are modelled using a Heaviside function, $H (\phi)$ as \cite{Gerlach2006Comparison}
\begin{eqnarray}
\rho( \phi) &=& {\rho _l}{{ H}(\phi) } + {\rho _g}( {1 -{ H}(\phi) }), \\
\mu( \phi) &=& {\mu _l}{{ H}(\phi) } + {\mu _g}( {1 -{ H}(\phi) }), 
\label{eq:viscosity level set}
\end{eqnarray}
where the subscripts `$l$' and `$g$' represent the liquid and vapor phases, respectively, and the Heaviside function $H(\phi)$ is given by \cite{Gerlach2006Comparison}

\begin{eqnarray}
H(\phi)= 
\left\{\begin{array}{l}
1~~~~~~~~~~~~~~~~~~~~~~~~~~~~~~~~~~~\mbox{if} ~~~~ \phi >  \epsilon,  \\
\frac{1}{2}+ \frac{\phi}{2 \epsilon} + \frac{1}{2 \pi} \left\{sin 
\left(\frac{\pi \phi}{\epsilon}\right)\right\} ~~~~\mbox{if} ~~~~  |\phi| \le  \epsilon, \\
0~~~~~~~~~~~~~~~~~~~~~~~~~~~~~~~~~~~\mbox{if}~~~~   \phi < - \epsilon. 
\end{array}\right.
\label{eq12}
\end{eqnarray}
The regularized Dirac delta distribution function, $\delta_s (\phi)$ is given by \cite{Gerlach2006Comparison}

\begin{eqnarray}
\delta_s (\phi) = 
\left\{\begin{array}{l}
\frac{1}{2 \epsilon} + \frac{1}{2 \epsilon} cos \left(\frac{\pi \phi}{\epsilon}\right) ~~~~\mbox{if} ~~~~  |\phi| \le  \epsilon, \\
0~~~~~~~~~~~~~~~~~~~~~~~~~~   {\rm otherwise}. 
\end{array}\right.
\label{eq13}
\end{eqnarray}

\subsection{Initial and boundary conditions}

The following initial and boundary conditions are implemented to perform the numerical simulations. At the start of the computation (at $t=0$), the fluids are assumed to be stationary with a fully developed flow imposed at the inlet of the tube $(z = 0)$ for $t \ge 0$. The boundary conditions at $z=0$ are given by
\begin{equation}
u(r) = 2 U_{avg} \left( 1 - {r^2 \over {R_0}^2} \right), ~ v(r) = 0, ~  T(r) = T_g, ~{\rm and} ~Y(r) = Y_g,
\end{equation}
where $T_g$  and $Y_g$  are the temperature and mass fractions of water vapor in the gas at the inlet of the tube.  At the tube wall ($r=R_0$), the no-slip and no-penetration conditions $(u = v = 0)$ are imposed. The Dirichlet conditions: $T= T_g$  and $Y= Y_g$ are imposed at the wall for temperature and species concentration. The symmetry boundary conditions for the flow variables, temperature and mass fraction are applied at $r=0$. At the outlet of the tube $(x=L)$, the Neumann boundary conditions for the velocity components, temperature and mass fraction are applied. The pressure at the outlet section is set to the atmospheric pressure. This acts as the Dirichlet boundary condition for the pressure at the tube outlet.

The various dimensionless numbers associated with this problem which will be used to present the results are the Reynolds number ($\Re=\rho_{g} U_{avg} D_0 / \mu_{g}$), Weber number ($\We = \rho_{g} U_{avg}^2 D_0 / \sigma$), Prandtl number ($\Pr = \mu_g  {c_{pg}}/ \kappa$), Schmidt number ($\Sc = \mu_g / \rho_g D_{\alpha g}$), Lewis number ($\Le = \kappa_v / \rho_g {c_{pg}} D_{\alpha g}$) and Jacob number ($\Ja = c_{pg} (T_v - T_{sat})/ h_{lg}$). The viscosity, density and conductivity ratios of fluids are given by $\mu_{r} = \mu_{l}/ \mu_{g}$, $\rho_{r} = \rho_{l}/ \rho_{g}$, $c_{{p},r} = {c_{pl}/ c_{pg}}$ and $\kappa_{lr} = \kappa_{l}/ \kappa_{g}$. The ratio of initial drop diameter $D_{drop}$ to that of tube diameter $({D_0})$ is termed as the aspect ratio ($a = {D_{drop}} / {D_0}$).
 
\subsection{Numerical method}

In the present study, the numerical simulations are performed in axisymmetric coordinates using the CLSVOF algorithm \cite{ Sussman2000Coupled,Gerlach2006Comparison}. A staggered MAC grid \cite{harlow1965numerical} is used as the basis for the spatial discretization. The grid dimensions in the axial and radial directions are equal. The flow is Newtonian and incompressible. The drop-liquids are mono-component liquids. We assume the thermophysical properties of both fluids to be constant. Simulations including phase change must accurately calculate the amount of mass transfer across the phase interface. Large density differences between the liquid and vapor phases can magnify small errors in mass transfer. In our method, the continuity equation is deployed to implement the mass transfer model through the use of a source term representing the mass transfer across the phase interface. The principle of modeling interface mass transfer using a source term in the continuity equation is similar to that of Juric and Tryggvason \cite{juric1998computations}. Our numerical methodology consists of three major steps, namely, the interface reconstruction, setting a phase change model, and solution of the Navier-Stokes equations. We begin our simulations by reconstructing the interface based on the initial configuration of the liquid droplet. The CLSVOF algorithm is deployed to capture the evolution/ advection of the interface. In the CLSVOF based calculations, the level set function is used to capture the tortuous interface and the volume of fluid function is used to conserve the mass. Our mass transfer model was described in Welch and Wilson \cite{Welch2000Volume} in connection with a Volume of Fluid (VOF) implementation.

The advection equation for the level set function is solved using the essentially non-oscillatory (ENO) scheme \cite{chang1996level}, while the advection of the VOF function is achieved through the geometric advection method. Once the location of the interface is identified in the computational domain, we use the embedded boundary method \cite{zhao2022boiling} to accurately capture the phase change process at the interface. The phase change algorithm consists of five sub-steps, namely, solving the species transport equation, estimating the interface temperature using an explicit relationship from Clausius-Clapeyron equation, solving the energy equation independently in both liquid and gaseous phases, calculating the vaporization rate using Eq. (\ref{jump2}) with the revised temperature field, and re-calculating the interface mass fraction using Eq. (\ref{eq8}). These five steps are repeated in every cycle to account for the phase change at the interface. The convective terms of the energy equation and species transport equation are discretized using the QUICK scheme of Leonard \cite{leonard1979stable}, while the diffusive terms are discretized using central difference approximation. Eq. (\ref{jump3}) is used to calculate the velocity jump conditions at the interface. We derive the pressure Poisson equation from the augmented continuity equation and the momentum equation, which is solved iteratively by preconditioned conjugate gradient scheme of Van der Vorst \cite{VanderVorst1992BICGSTAB}. The advection terms of the momentum equations are discretized using the ENO scheme \cite{chang1996level}, while the diffusive terms are discretized using the central difference scheme. The detailed formulation and numerical method utilized in this study can also be found in Ref. \cite{pal2023clsvof}.

\begin{figure}[h]
\centering
(a) \\
\includegraphics[width=0.35\textwidth]{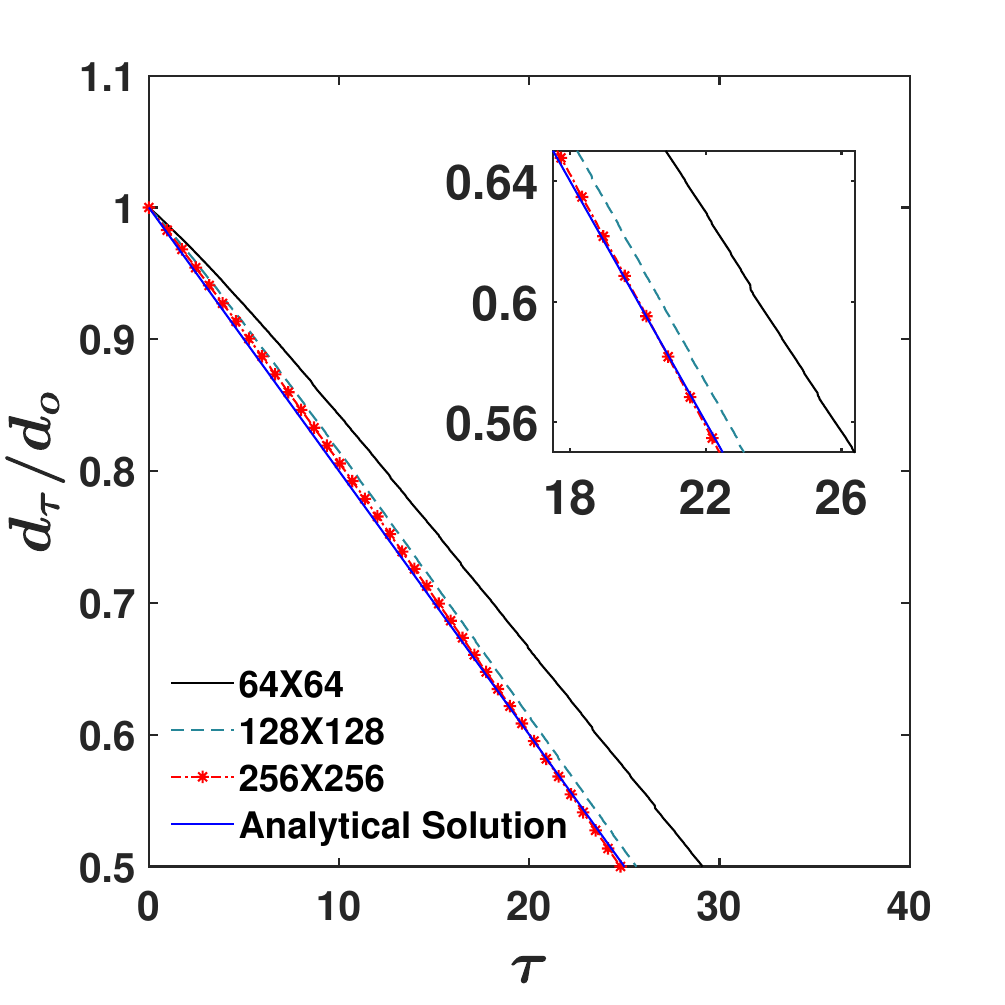} \\
(b) \\
\includegraphics[width=0.35\textwidth]{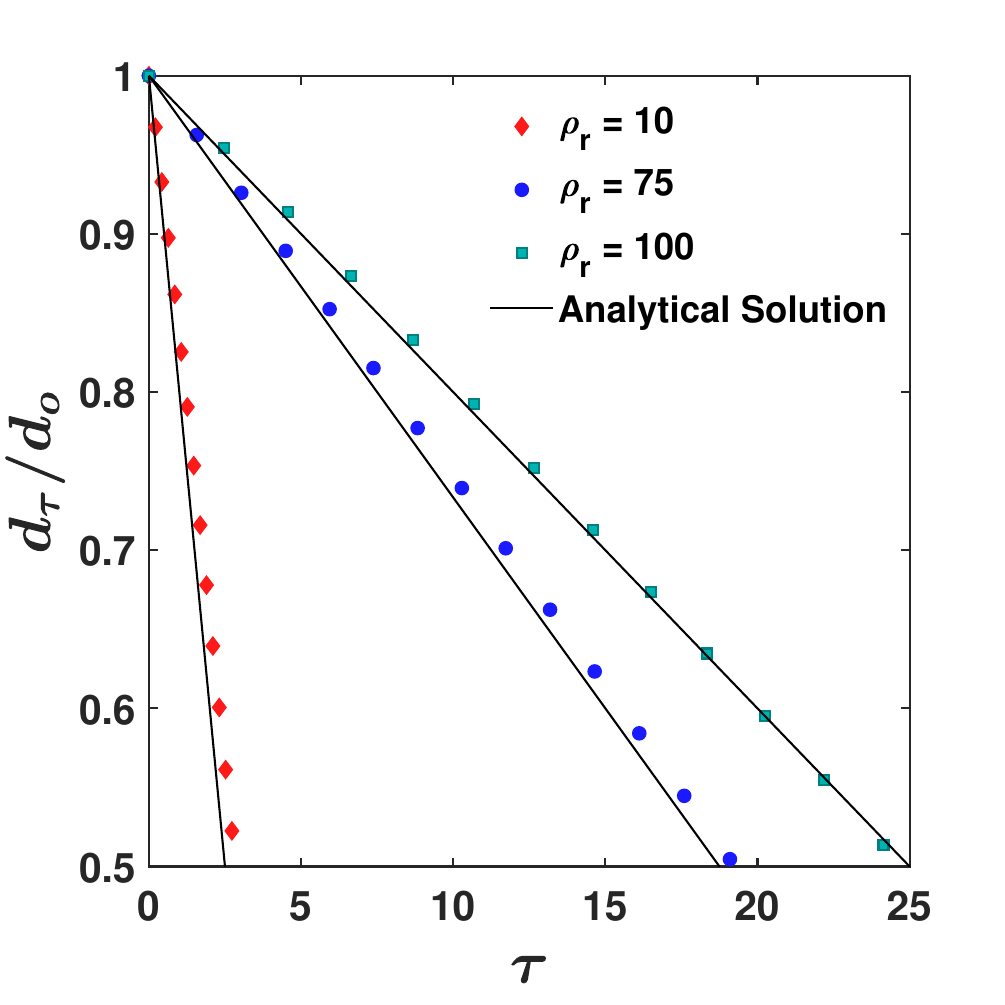} 
\caption{Evaporation of a static drop in quiescent medium. (a) Grid convergence test for an evaporating drop with $\rho_r = 100$. (b) Temporal evolution of the normalized drop diameter for $\rho_r = 10$, 75 and 100. The rest of the dimensionless parameters are $Re=25$ ,$We=0.1$ and $\mu_r = 50$.}
\label{fig2}
\end{figure}

\begin{figure}[h]
\centering 
(a) \\
\includegraphics[width=0.35\textwidth]{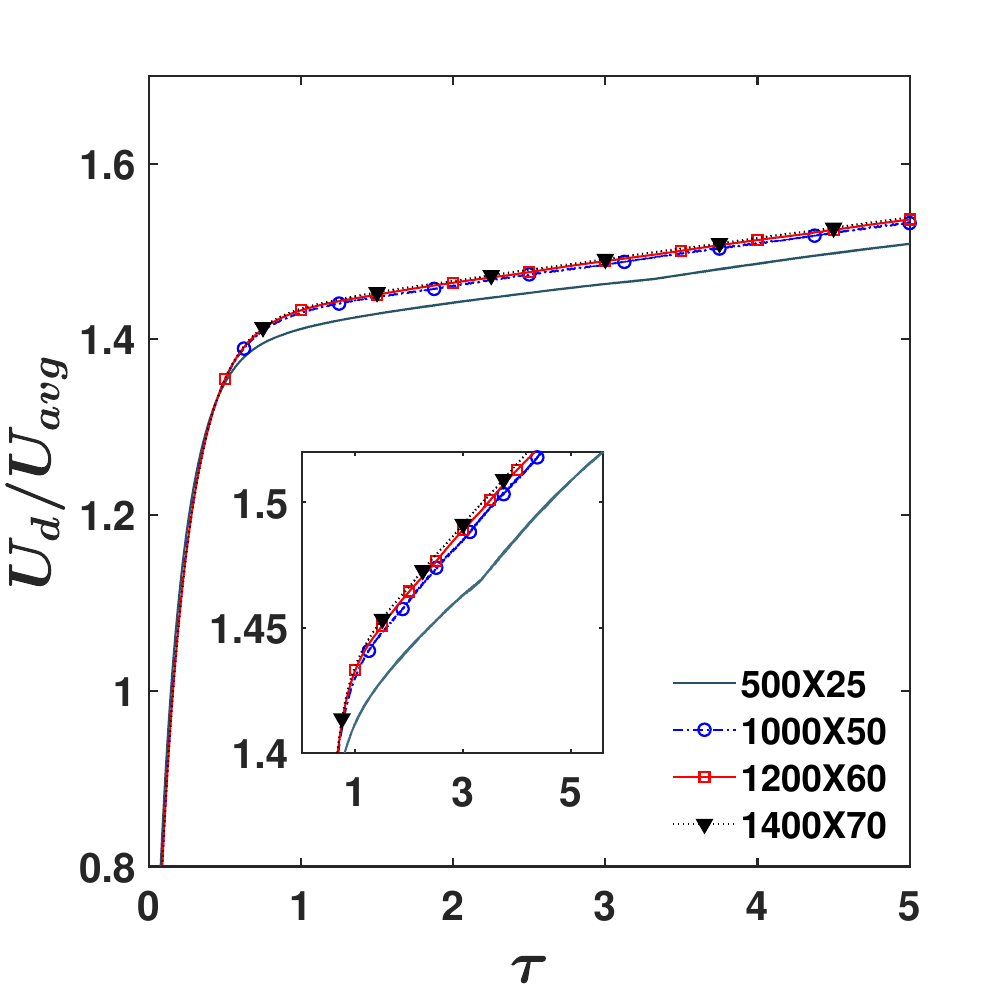} \\
(b) \\
\includegraphics[width=0.35\textwidth]{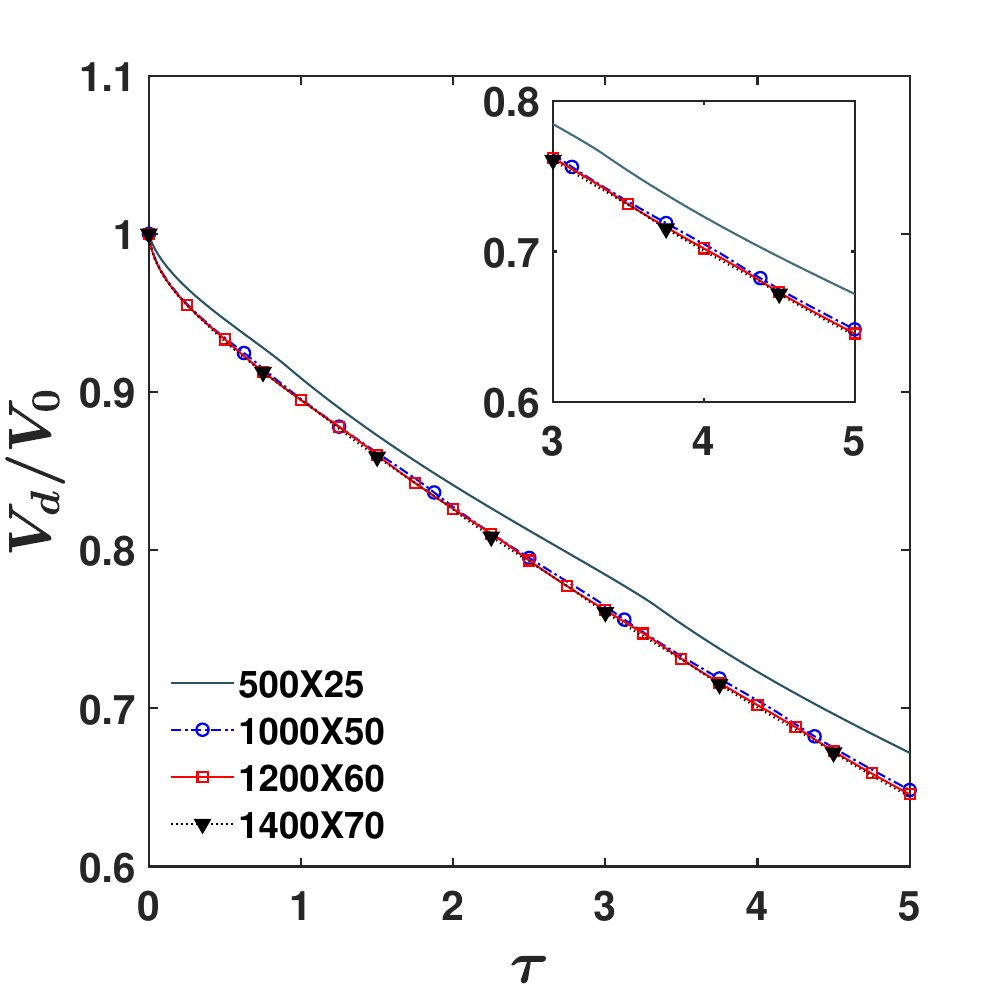} 
\caption{Grid convergence test for an evaporating drop (with aspect ratio, $a = 0.7$) migrating in a tube. The drop and gas phases are initialized at 313 K and 393 K, respectively. The temporal variations of (a) the normalised migration velocity of the drop $(U_d/U_{avg})$ and (b) normalised drop volume $(V_d/V_0)$ obtained using four different grids. The rest of the dimensionless parameters are $\Re=40$, $\rho_r = 5$, $\mu_r = 10$, $\We = 80$, $\Le = 1$, $\Pr = 3.87$ and $\Ja = 0.0083$.}
\label{fig3}
\end{figure}

\subsection{Validation}
We present the quantitative validation of our phase change model in this section for a benchmark case. We validated our solver for constant mass flux-based evaporation case as suggested in the literature \cite{scapin2020volume}. The analytical solution for the normalized instantaneous diameter of the drop for this case is given by
\begin{equation}
{d_t \over d_0} = 1 - {2m_0 \over d_o \rho_l}t.
\end{equation}
A static liquid drop of diameter $d_0$ is kept at the center of a square domain of side $4d_0$. The interface is given an arbitrary velocity such that $Re=25$ and $We=0.1$. We first conduct a grid convergence test for a drop with $\rho_r = 100$ as shown in figure \ref{fig2}a. Here, $\tau = {2m_{0}t / d_o \rho_l}$ is the dimensionless time. However, it is to be noted that, in the following, while studying the migrating drop, the non-dimensional time, $\tau$, is defined as $ \tau =  {tU_{avg} /D}$. Three different grid sizes are considered, namely $64 \times 64$, $128 \times 128$ and $256 \times 256$. It is evident that the solution corresponding to $256 \times 256$ grid agrees reasonably well with the analytical solution. Therefore, the grid with $256 \times 256$ grid points is used in further validation cases pertaining to the evaporation of a static liquid drop. In figure \ref{fig2}b, we compare the numerical solution for three different values of density ratios $\rho_r = 10$, 75 and 100 with corresponding analytical solutions. The accurate match establishes the ability of our solver to capture the evolution of the interface without the formation of spurious currents. Furthermore, to validate the $d^2$ law of evaporation, several exercises based solely on temperature gradient-induced evaporation were conducted, which are detailed in Ref. \cite{Dubey}.

\begin{figure*}[h]
\centering 
 \hspace{-0.3cm} (a) \hspace{2.2cm} (b)  \hspace{2.2cm} (c) \hspace{2.2cm} (d)\\
\includegraphics[width=0.7\textwidth]{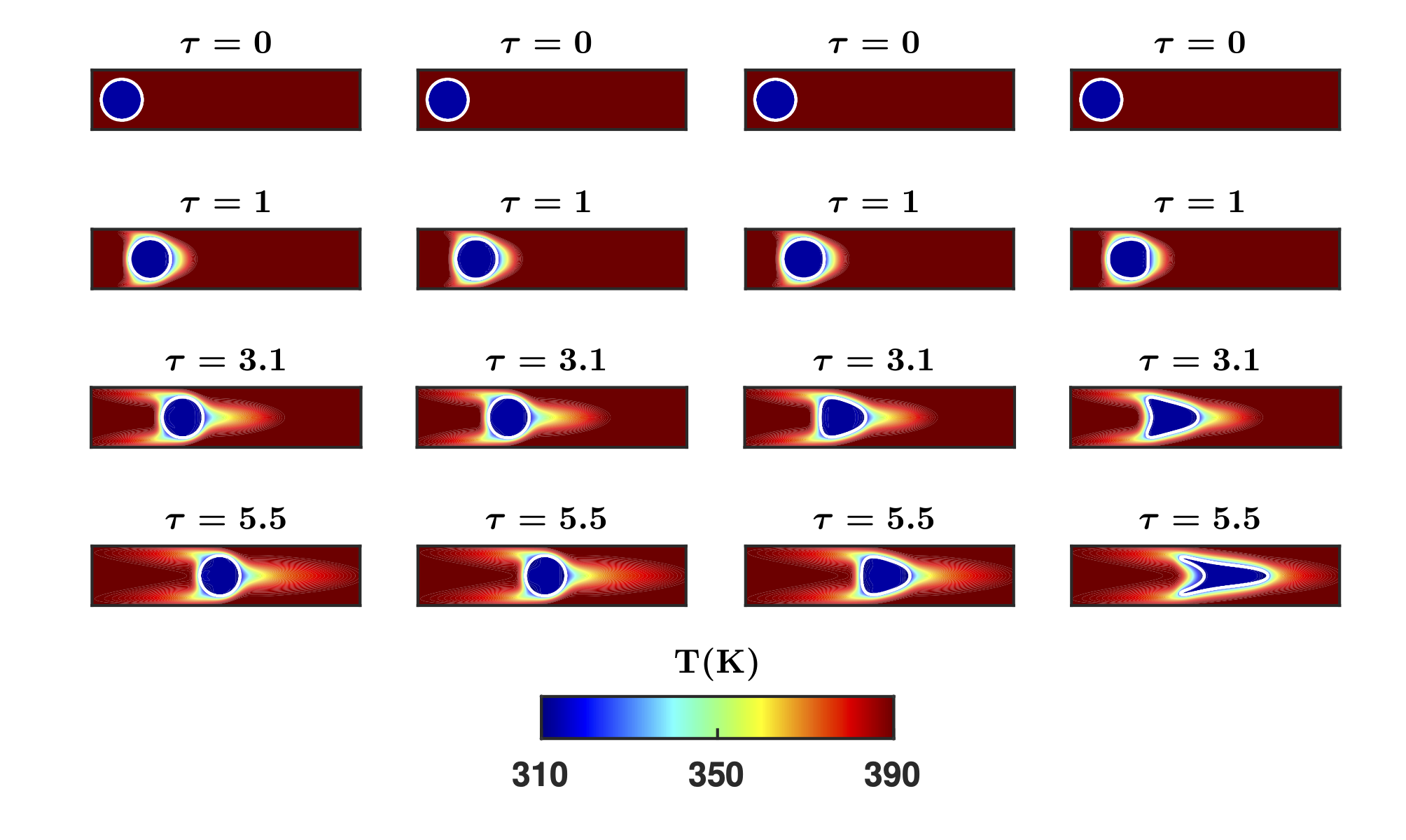} 
\caption{Temporal evolution of drop profile and temperature field for (a) $\We = 0.1$, (b) $\We = 1$, (c) $\We = 10$ and (d) $\We = 100$ for a drop with  $a = 0.7$. The rest of the dimensionless parameters are $\rho_r = 5$, $\mu_r = 10$, $\Le = 1$, $\Re = 40$, $\Pr = 3.87$ and $\Ja = 0.0083$.}
\label{fig4}
\end{figure*}

Further,  a grid convergence test is conducted for an evaporating drop migrating in a tube (Figure \ref{fig1}) for a typical set of parameters: $a=0.7$, $\Re=40$, $\rho_r = 5$, $\mu_r = 10$, $\We = 80$, $\Le = 1$, $\Pr = 3.87$, and $\Ja = 0.0083$. Figure \ref{fig3}(a) and (b) show the temporal variations of the normalised migration velocity of the drop $(U_d/U_{avg})$ and normalised drop volume $(V_d/V_0)$ obtained using four different grids, namely, $500 \times 25$, $1000 \times 50$, $1200 \times 60$, $1400 \times 70$. It can be seen that the results obtained from grids $1000 \times 50$, $1200 \times 60$, and $1400 \times 70$ are found to be identical, and thus, $1000 \times 50$ is chosen as the optimum grid size for generating the remaining results. Notably, while Nath et al. \cite{nath2017migration} observed a steady-state migration velocity for a non-evaporating drop, our study observed a monotonic increase in drop velocity as it migrated through the tube, as depicted in Figure \ref{fig3}(a).

\section{Results and discussion} \label{sec:dis}

Our study examines how a drop migrating inside a tube behaves as it evaporates. Specifically, we focus on how the migration and the resulting deformation affect the evaporation of the drop and vice versa. To measure the heat and mass transfer during this process, we use the Nusselt and Sherwood numbers. The Nusselt number, $\Nu (=h_{avg} D_{eq} / k_g)$ is the ratio of the convective heat transfer coefficient $(h)$ to the thermal conductivity of the gas $(k_g)$. The Sherwood number is defined as $\Sh=\dot m D_{eq} / \cal A$. Here, $h_{avg}$ denotes the average heat transfer coefficient calculated using the cumulative average over all the computational cells at the drop periphery; ${\dot m}$ is the vaporization rate, $D_{eq}$ denotes the instantaneous equivalent spherical diameter of the drop and ${\cal A} ={\rho_g D_{\alpha g} \ln {(1-Y_{\infty})/(1-Y_{\Gamma})}}$ \cite{zhao2022boiling}, that represents the conductive mass transfer taking place due to diffusion arising from the concentration gradient at the drop periphery. The non-dimensional time, $\tau$, is obtained from the average velocity of imposed flow as the velocity scale and tube diameter as the length scale, which is given by $\tau =  {tU_{avg} /D}$. We found (not shown) that $\Nu$ and $\Sh$ achieve a steady value after an initial transient period, which is consistent with the results of Ref. \cite{rantz1952evaporation} despite using a uniform airflow. Next, we investigate the effect of the Weber number $(\We)$, Jacob number $(\Ja)$, and aspect ratio $(a = {D_{drop}} / {D_0})$ on the dynamics of an evaporating drop migrating in a tube. We also observed that within the range of parameters considered, the heat and mass transfer coefficients are independent of the density ratio ($\rho_r$) and viscosity ratio ($\mu_r$). Thus, we select $\rho_r = 5$ and $\mu_r = 10$ in our study. Finally, we establish a scaling law between the dimensionless parameters.

\subsection{Effect of Weber number}\label{subsec:We}

\begin{figure}[h]
\centering 
\includegraphics[width=0.35\textwidth]{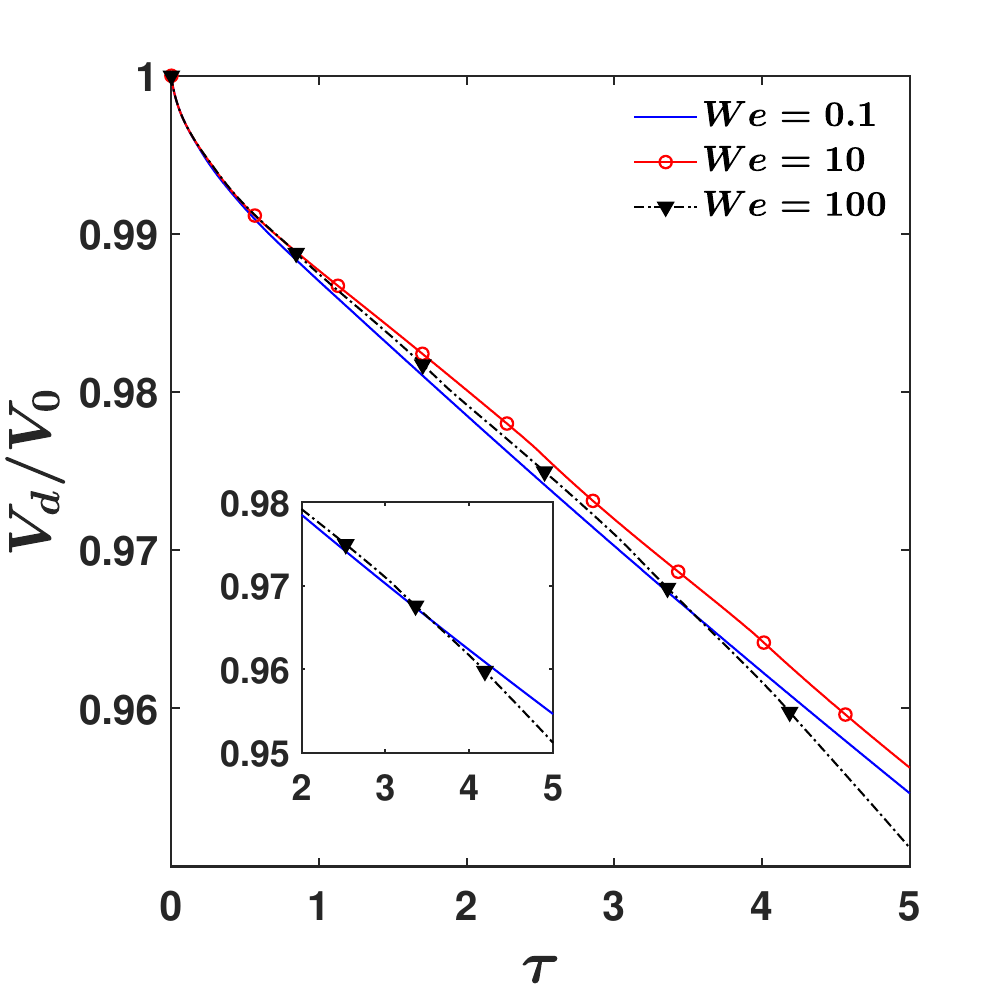} 
\caption{Temporal variation of the normalised instantaneous volume of the drop $(V_d/V_0)$ for different values of $\We$. The rest of the dimensionless parameters are $a = 0.7$, $\rho_r = 5$, $\mu_r = 10$, $\Le = 1$, $\Re = 40$, $\Pr = 3.87$ and $\Ja = 0.0083$.}
\label{fig5}
\end{figure}

\begin{figure*}[h]
\centering 
 \hspace{0.2cm} (a) \hspace{3.0cm} (b)  \hspace{3.0cm} (c) \\
\includegraphics[width=0.7\textwidth]{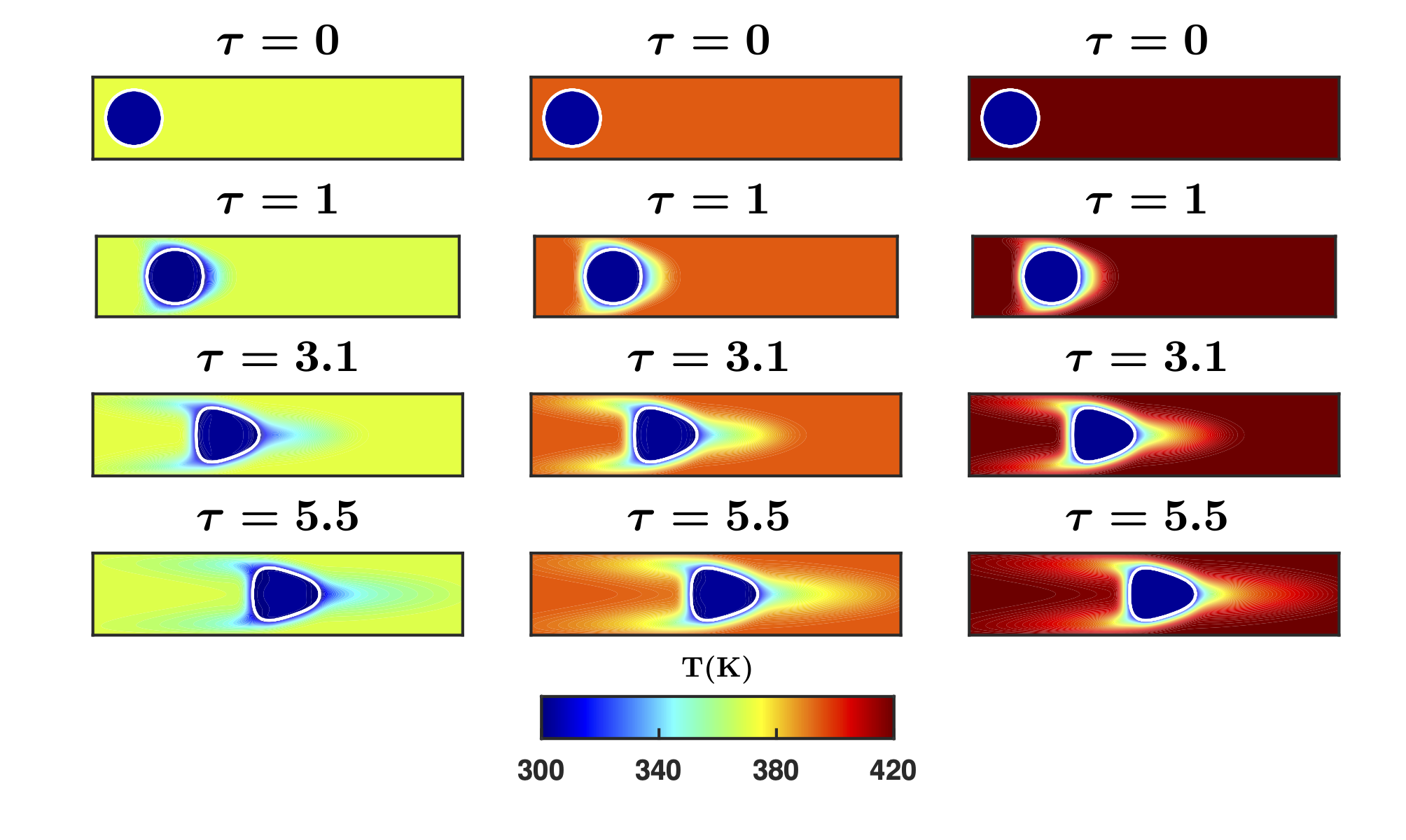} 
\caption{Temporal evolution of drop profile and temperature field for (a) $\Ja = 0.001$, (b) $\Ja = 0.01$ and (c) $\Ja = 0.02$ for a drop with $a = 0.7$. The rest of the dimensionless parameters are $\rho_r = 5$, $\mu_r = 10$, $\Le = 1$, $\Re = 40$, $\Pr = 3.87$ and $\We = 10$.}  
\label{fig6}
\end{figure*}

The behavior of a liquid drop undergoing deformation is determined by the relative strengths of aerodynamic drag and surface tension force at the drop periphery, which is quantified by the Weber number ($\We$). For a neutrally buoyant migrating drop, the capillary number ($\Ca$) is used as a counterpart to the Weber number, as demonstrated in previous studies by Nath et al. \cite{nath2017migration} on non-evaporating drops migrating in tubes. The Weber number ($\We = \rho_{g} U_{avg}^2 D_0 / \sigma$) characterised the competition between the inertial force due to the imposed flow inside the pipe and the surface tension force. A higher value of $\We$ indicates a dominance of the inertia force over the surface tension force, which promotes the deformation of the drop as it migrates inside the tube, as depicted in figure \ref{fig4}(a-c). To investigate the evaporation of deforming drops, we varied the Weber number from 0.1 to 100. To vary the Weber number $(We)$, we change the surface tension value $(\sigma)$ while keeping the Reynolds number $(Re)$ constant. Figure \ref{fig4}(a-c) shows the temporal evolution of the drop profile and the thermal boundary layer near the drop periphery for three different Weber numbers: 0.1, 10, and 100. When $\We = 0.1$, the drop retains its spherical shape during migration (Figure \ref{fig4}a). The convective gas flow and Stefan flows resulting from surface evaporation affect the temperature and concentration fields around the drop. For intermediate ($\We = 10$; Figure \ref{fig4}b) and large ($\We = 100$; Figure \ref{fig4}c) values of $\We$, the aerodynamic drag is strong enough to deform the drop. At $\We = 10$, the drop takes an oblate shape with zero curvature at the leading edge, while for $\We = 100$, the pressure distribution around the drop governs the deformation dynamics. The pressure is high at the leading edge due to air stagnation, lower between the drop periphery and the tube wall, and regains its value at the trailing edge. At $\tau = 1$, the trailing edge has zero curvature, and subsequently, the curvature at the leading edge changes sign as the drop migrates further.

Figure \ref{fig5} depicts the temporal variation of the instantaneous volume of a drop $(V_d)$ with its initial volume $(V_0)$ is shown for various $\We$ values. An interesting behavior is observed during the initial transient period, where the rate of evaporation increases with deformation, as reported by Nazafian et al. \cite{najafian2023numerical} in their investigation of the evaporation of a falling drop in a wide channel with no boundary effects. Surprisingly, the drop with $\We = 0.1$ evaporates faster than the one with $\We = 10$, despite the latter having a larger surface area. The cause of this anomaly is attributed to the way the transport coefficients vary over the periphery of the drop. When the free stream flow approaches the drop, it first accelerates in the gap between the drop periphery and tube wall and subsequently decelerates. In the case of $\We = 10$, the deceleration occurs over a larger length because of the liquid drop's tapered shape, compared to the spherical drop of $\We = 0.1$. However, for $\We = 100$, although the evaporation rate is initially lower, it eventually becomes greater than that of a spherical drop because the increase in surface area counteracts the effect of low gradients in the drop's wake. In conclusion, the peripheral gap and surface area govern the evaporation rate for a drop evaporating in a tube.

\subsection{Effect of superheat}\label{subsec:Ja}

\begin{figure}[h]
\centering 
\includegraphics[width=0.4\textwidth]{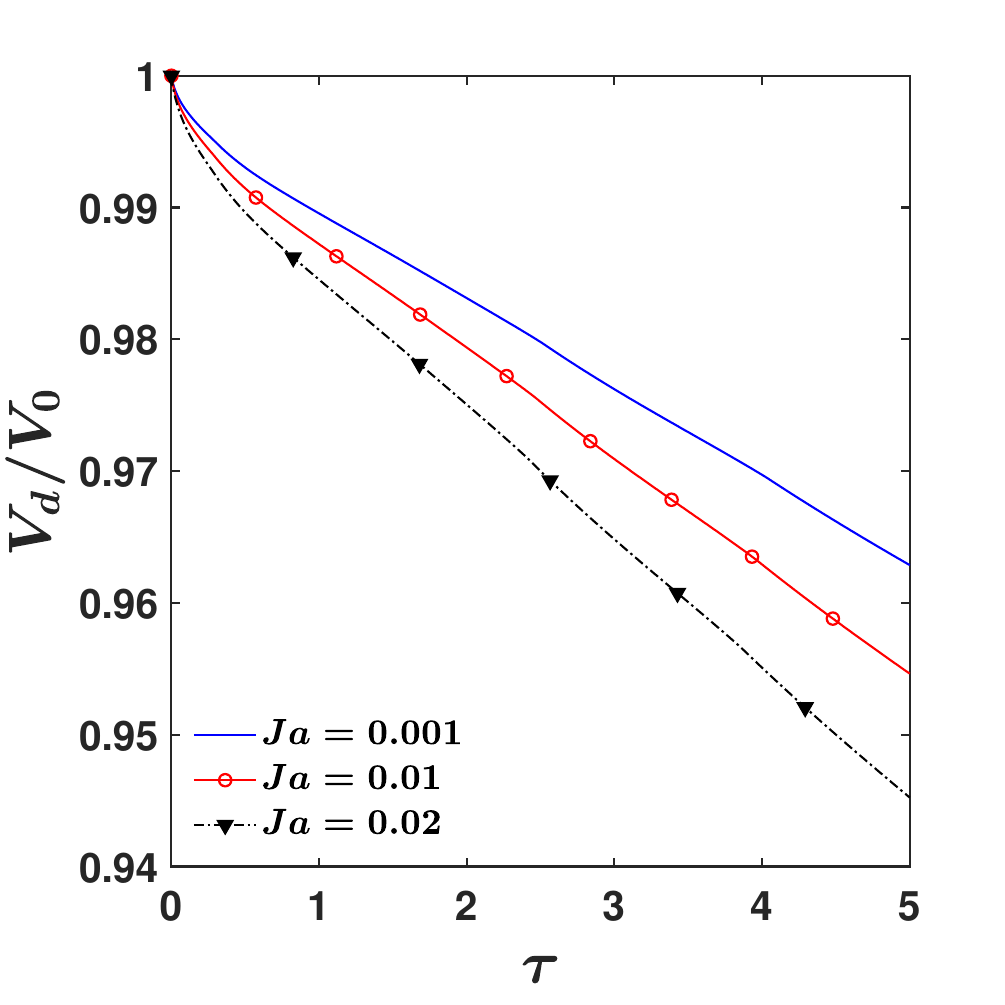} 
\caption{Temporal variation of the normalised instantaneous volume of the drop $(V_d/V_0)$ for different values of $\Ja$. The rest of the dimensionless parameters are $a = 0.7$, $\rho_r = 5$, $\mu_r = 10$, $\Le = 1$, $\Re = 40$, $\Pr = 3.87$ and $\We = 10$.}
\label{fig7}
\end{figure}

\begin{figure*}[h]
\centering 
 \hspace{-0.5cm} (a) \hspace{2.2cm} (b)  \hspace{2.2cm} (c) \hspace{2.2cm} (d)\\
\includegraphics[width=0.7\textwidth]{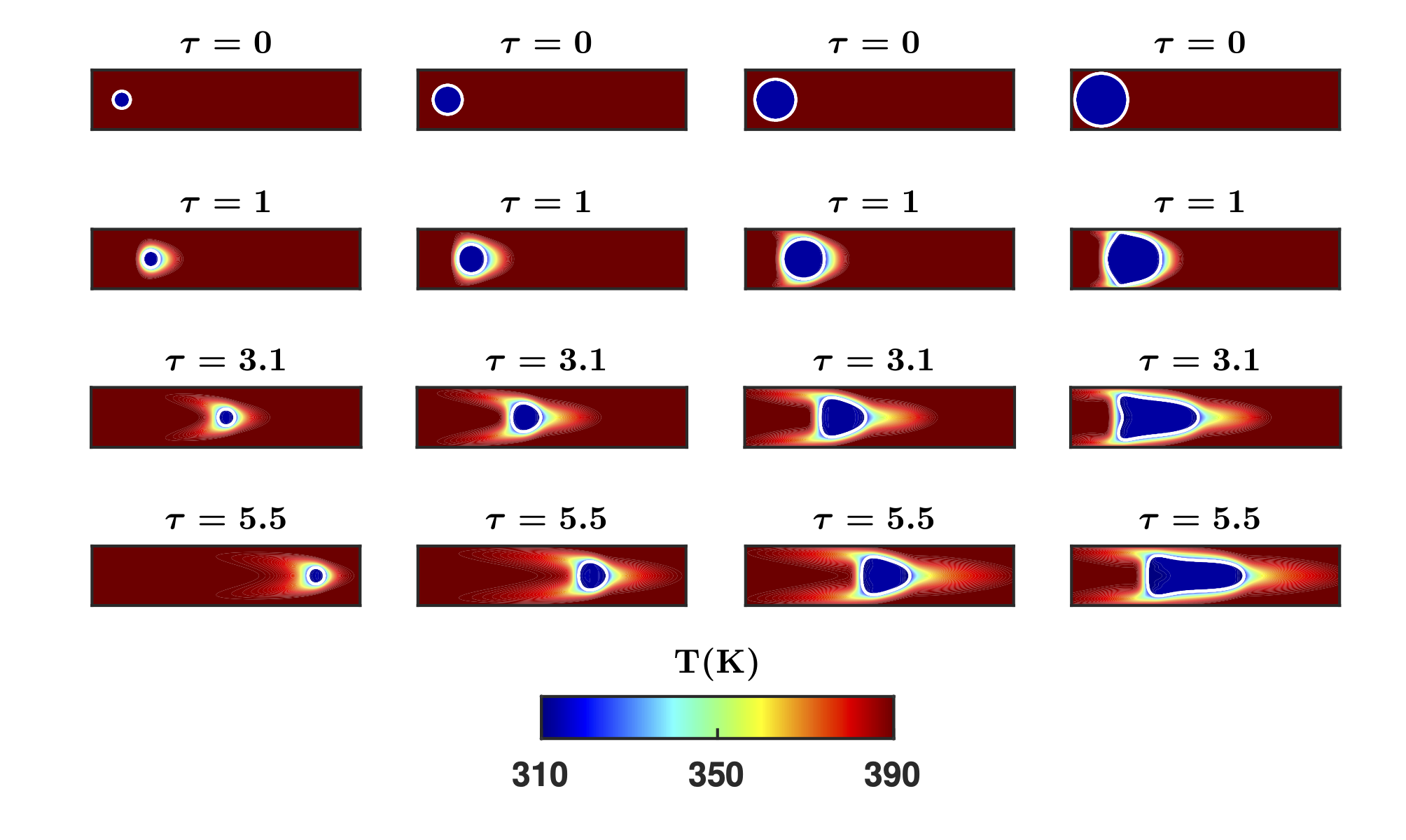} 
\vspace{-0.5cm} 
\caption{Temporal evolution of the morphology of drop and temperature field for (a) $a = 0.3$, (b) $a = 0.5$, (c) $a = 0.7$ and (d) $a = 0.9$. The rest of the dimensionless parameters are $\rho_r = 5$, $\mu_r = 10$, $\Le = 1$, $\Re = 40$, $\Pr = 3.87$, $\Ja = 0.0083$ and $\We = 10$.}
\label{fig8}
\end{figure*}

\begin{figure*}[h]
\centering 
 \hspace{-0.5cm} (a) \hspace{2.2cm} (b)  \hspace{2.2cm} (c) \hspace{2.2cm} (d)\\
\includegraphics[width=0.7\textwidth]{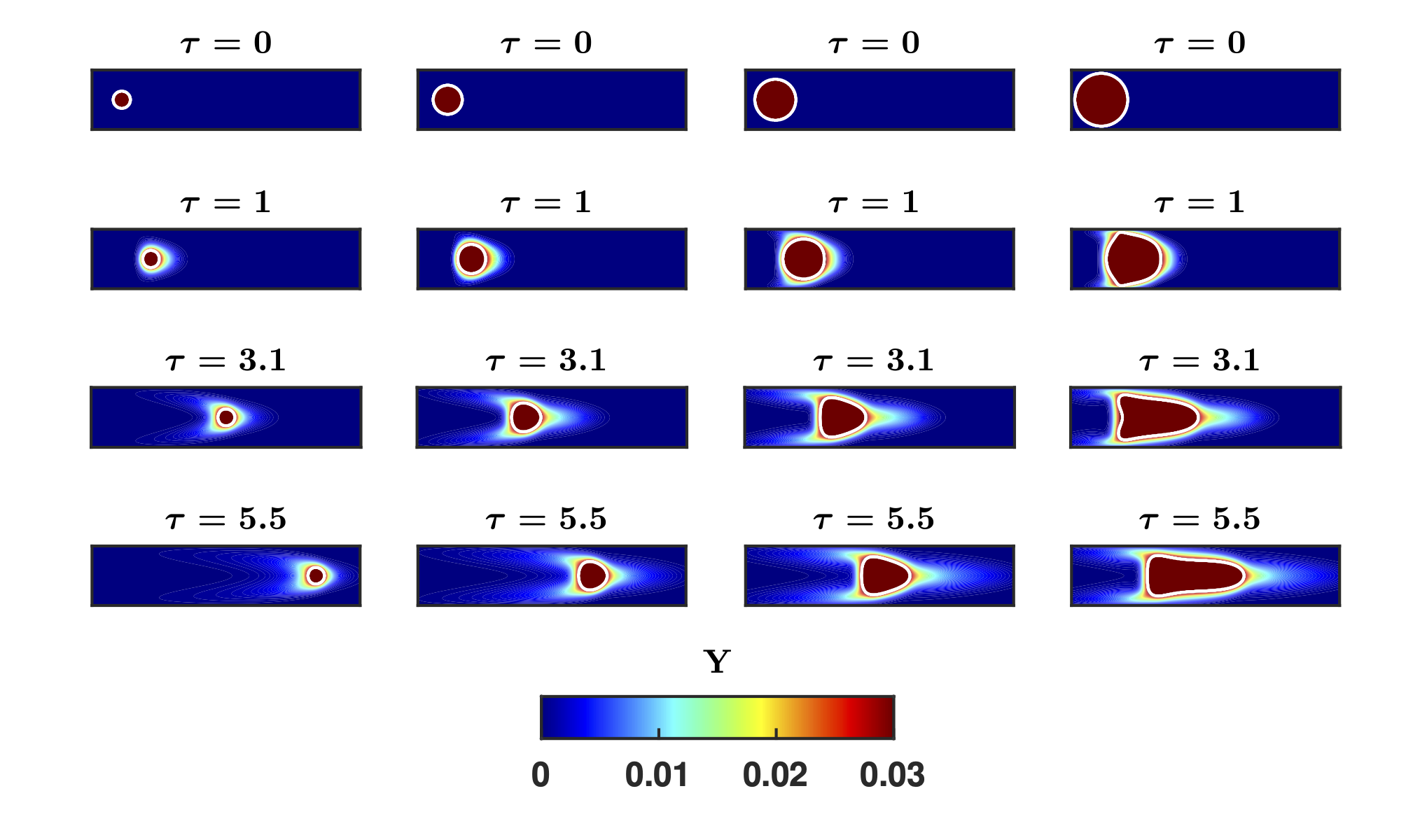}
\vspace{-0.5cm}  
\caption{Temporal evolution of the morphology of drop and mass fraction field for (a) $a = 0.3$, (b) $a = 0.5$, (c) $a = 0.7$ and (d) $a = 0.9$. The rest of the dimensionless parameters are $\rho_r = 5$, $\mu_r = 10$, $\Le = 1$, $\Re = 40$, $\Pr = 3.87$, $\Ja = 0.0083$ and $\We = 10$.}
\label{fig9}
\end{figure*}

\begin{figure*}[h]
\centering 
(a) \hspace{5.8cm} (b) \\
\includegraphics[width=0.35\textwidth]{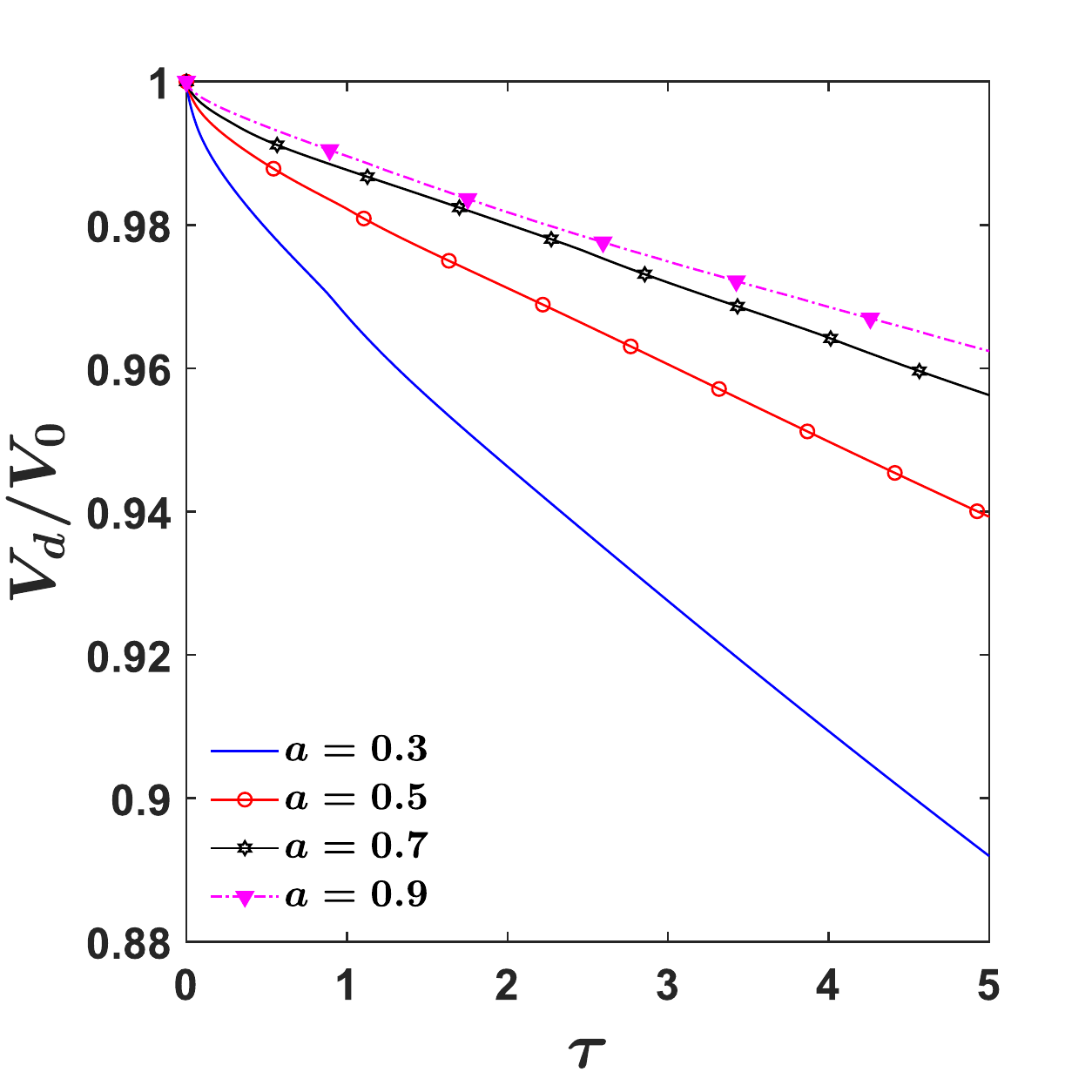} \hspace{0.2cm}
\includegraphics[width=0.35\textwidth]{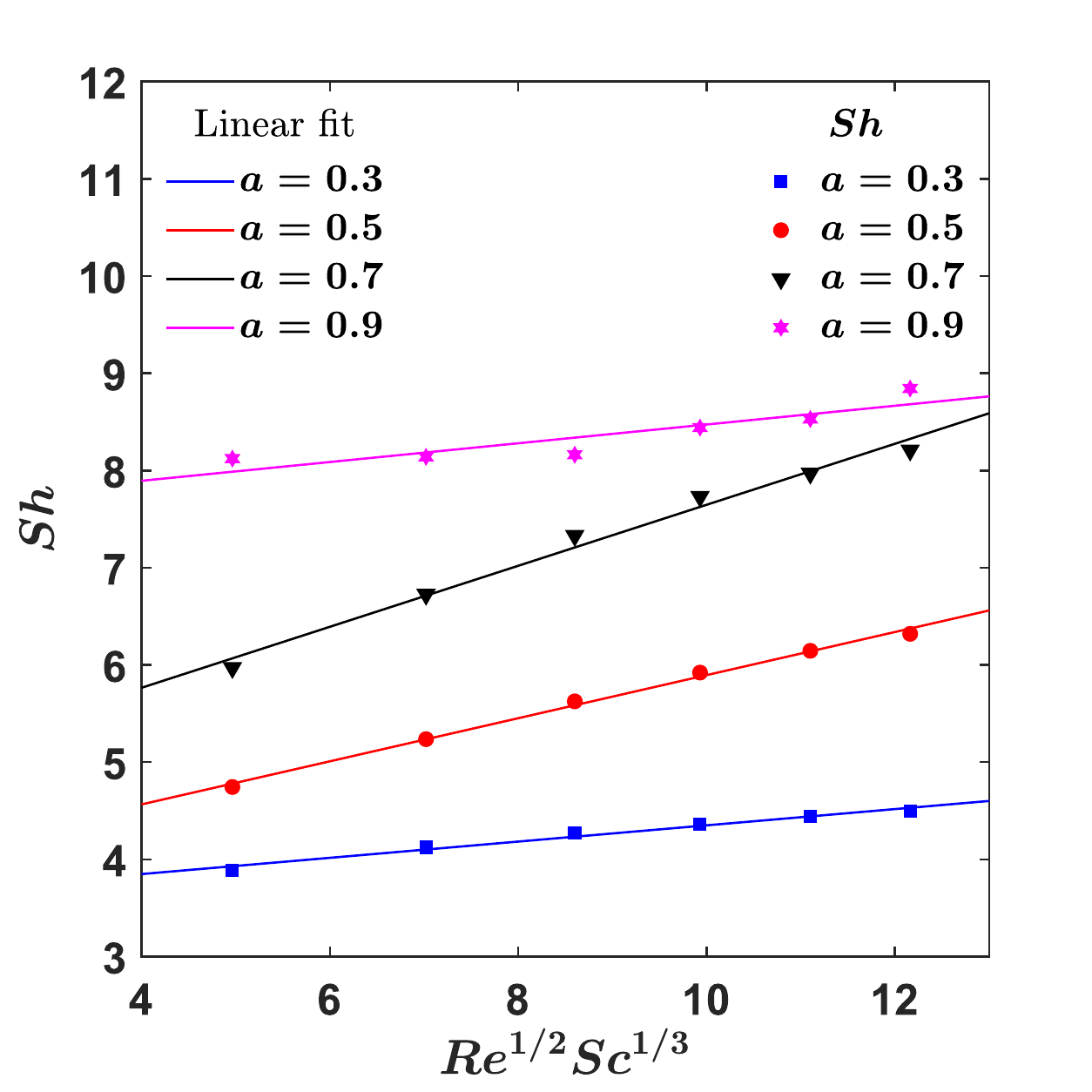}
\caption{(a) Temporal variation of the normalised instantaneous volume of the drop $(V_d/V_0)$ for different values of the aspect ratio $(a)$ for $\Re = 40$. (b) Variation of $ Sh $ with $ Re^{1/2}Sc^{1/3} $ for different values of $a$. The rest of the dimensionless parameters are $\rho_r = 5$, $\mu_r = 10$, $\Le = 1$, $\Pr = 3.87$, $\We = 10$ and $\Ja = 0.0083$.}
\label{fig10}
\end{figure*}

The superheat is defined as the excess temperature over the saturation temperature at a given ambient pressure and is characterized by the Jacob number $(\Ja)$. The higher superheat leads to  more significant temperature gradient at the drop periphery, resulting in rapid evaporation. To investigate its effect, we considered three gas temperatures corresponding to $\Ja = 0.001$, 0.01, and 0.02 while maintaining $\We = 10$ for the simulations without a re-entrant cavity. Figure \ref{fig6}(a-c) display the temporal evolution of the drop and the thermal boundary layer near the drop periphery for each case. Although an increase in the superheat results in rapid evaporation, which intensifies the associated Stefan flow, we observe the same development of thermal and concentration boundary layer in all three cases and an identical drop profile. Our findings support the results of Ref. \cite{Deng1992TwoDimensional}, which suggest that the evaporation rate does not significantly affect the deformation dynamics of an evaporating drop. However, the evaporation rate is governed by the deformation of the drop, as discussed in section \ref{subsec:We}. The temporal evolution of the instantaneous normalized drop volume with varying $\Ja$ values is depicted in Figure \ref{fig7}. Surprisingly, despite the proximity between the drop periphery and the tube walls, where the latter are subject to Dirichlet boundary conditions for temperature, only 4\% - 7\% of the liquid volume evaporated, indicating a high residence time for these small liquid drops. These findings support the results reported in Ref. \cite{pedraza2018numerical}, which showed only 1.2\% mass loss for a 1 mm diameter drop.

\subsection{Effect of aspect ratio}\label{subsec:a}

Finally, we examine the effect of aspect ratio ($a = {D_{drop}} / {D_0}$) on heat and mass transfer by considering three different values of aspect ratio $a=0.3$, 0.7, and 0.9 for $\We=10$. To examine the effects of the aspect ratio, we changed the initial drop diameter while keeping the tube diameter constant. The temporal evolution of the temperature field and mass fraction field around the drop profile for the three cases are presented in Figures \ref{fig8} and \ref{fig9}, respectively. Temporal variations of the normalised instantaneous volume of the drop $(V_d/V_0)$ for different values of $a$ are depicted in Figure \ref{fig10}(a). It can be seen that increasing $a$ decreases the evaporation rate. We observe that the variation of Sherwood number ($\Sh$) with $\Re^{1/2}\Sc^{1/3}$ is linear for all three values of $a$ considered in the present study, as shown in Figure \ref{fig10}(b). This linear variation has been reported in earlier studies \cite{Reutzsch2020Consistent, rantz1952evaporation} based on evaporation brought about by uniform airflow without considering any boundary effects. Surprisingly, in the case of an evaporating liquid drop migrating in a tube, the linear variation holds despite the deformation. A thorough scrutiny of figure \ref{fig10}(b) reveals that for a given $\Re$, the value of $\Sh$ increases with an increase in aspect ratio due to the increased blockage in the path of flowing gas, causing it to accelerate rapidly through the reduced gap between the drop periphery and tube wall, thereby enhancing evaporation. Interestingly, we observe a non-monotonic rate of increase in $\Sh$ between the cases of $a=0.3$, 0.7, and 0.9. The drop with an aspect ratio of $a=0.3$ shows no deformation, whereas the intermediate ($a=0.7$) and large drops ($a=0.9$) deform as they migrate through the tube and eventually attain a steady-state shape for the range of $\Re$ considered. The larger drop undergoes higher deformation; therefore, the characteristic length for the drop responsible for heat and mass transfer decreases. Thus, the rate of increase of $\Sh$ with $\Re^{1/2}\Sc^{1/3}$ is higher in the case of $a=0.7$ compared to $a=0.9$. For $\We=10$, if the Reynolds number is further increased, the drop will not assume a steady-state profile, and therefore the $\Sh$ number will keep decreasing.

\section{Conclusions} \label{sec:conc}

We numerically examine the dynamics of an evaporating liquid drop migrating in a tube using the coupled level set and volume of fluid (CLSVOF) method. This study aims to gain insights into the evaporation mechanisms of liquid drops migrating in a narrow tube. Depending on the surface tension of the liquid, the drop can retain its shape or deform, leading to disintegration. We have developed a model to solve energy and species transport equations and validated it for several benchmark problems, such as – temperature-gradient-driven evaporation, concentration gradient-driven evaporation and coupled evaporation. It is then applied to a drop migrating in a Hagen - Poiseuille flows through a narrow tube. The effects of density ratio, viscosity ratio, degree of superheat and Weber number are investigated. We investigate the effect of $\rho_{r}$ and $\mu_{r}$ on $\Nu$ and $\Sh$, which is found to be insignificant for the range of parameters considered in this study. These findings enabled us to consider a feasible model for studying nuances of evaporation of a migrating drop with superimposed deformation effects. Furthermore, we conducted an extensive parametric study for a situation with a unit Lewis number. The Reynolds number and Weber number were varied to compare the impact of drop deformation on heat and mass transfer coefficients. We found that the deformation of a drop to oblate shapes reduces $\Nu$ and $\Sh$. An increase in aspect ratio increased the blockage in the tube, thereby increasing gas acceleration at the periphery of the drop, thus causing increased evaporation and higher values of $\Nu$ and $\Sh$. 

\section*{Acknowledgment}
G.B. acknowledges the J. C. Bose National Fellowship of SERB, Government of India (Grant No.\ JBR/2021/000042). K. C. S. thanks the Science \& Engineering Research Board, India (Grant No.\ CRG/2020/000507) for financial support. We acknowledge the help received from Param Sanganak Computing system of IIT Kanpur.

\end{document}